\documentclass[prl,12pt,showpacs,floatfix,preprintnumbers,amsmath,amssymb]{revtex4}

\usepackage{graphicx}
\usepackage{dcolumn}
\usepackage{bm}

\usepackage{color}

\begin{document}


\title{Many-body theory for angular resolved photoelectron spectra of metal clusters}
\author{Andrey V. Solov'yov}
\homepage{http://www.fias.uni-frankfurt.de/mbm/}
\email[Email address: ]{solovyov@fias.uni-frankfurt.de}
\altaffiliation[On leave from: ] {A.F. Ioffe Physical-Technical Institute,
St Petersburg, Russia}
\affiliation{Frankfurt Institute for Advanced Studies, Ruth-Moufang-Str. 1, D-60438 Frankfurt am Main, Germany}

\author{Roman G. Polozkov and Vadim K. Ivanov}
\homepage{http://www.physics.spbstu.ru}
\affiliation{St. Petersburg State Polytechnical University, 
Politekhnicheskaya 29, Saint-Petersburg 195251, Russia}
\email[Email address: ]{ivanov@tuexph.stu.neva.ru}

\date{\today}

\begin{abstract}
Angular resolved photoelectron spectra of metal clusters have been experimentally measured for the first time
only recently \cite{Issendorf}.
These measurements have been performed systematically for sodium clusters in a broad range of cluster sizes.
This work attracted a lot of attention and was reported practically at all major international cluster conferences 
because it revealed a very  non-trivial behavior of the angular anisotropy parameter with respect to photon energy and
provided a method for probing the angular momentum character of the valence orbitals of free nanoclusters.
Initial attempts to explain these observations within single particle approximations fail completely  \cite{Issendorf}. 
In this Letter we present a consistent many-body theory for the description of angular resolved photoelectron 
spectra of metal clusters. Jellium model formalism is employed.  
Our calculations demonstrate the dominant role of the many-body effects in the formation
of angular distributions of photoelectrons emitted from sodium clusters and are in a good agreement with
experimental data reported in \cite{Issendorf}. The concrete comparison of theory and experiment has been performed for the
photoionization of $Na_7^{-}$ and  $Na_{19}^{-}$ anions being characterized by the entirely closed shells of delocalized
electrons.

\end{abstract}

\maketitle


The problem of photoionization of metal clusters has been studied intensively during the
last two decades, see e.g. \cite{Solovyov} and references therein.  During this period 
a number of important achievements have been
made. Thus, the plasmon resonances have been observed in small metal clusters 
of nanometer size for different materials, e.g. Na, K,  Mg,
in fullerenes and in many other cases of study, for references see \cite{Solovyov}. 
The structure of these plasmon resonances
have been studied in detail. It was established that  patterns of plasmon resonances in many
cases are determined by the deformation parameter of the cluster. This result is 
very important because it provides a method of monitoring the cluster size and its
shape. A comprehensive review of the results in this field can be found in \cite{Solovyov_IJMPB}.

The detail geometrical structure of mass selected
clusters has been also studied by means of photoelectron spectroscopy and the density functional theory (DFT).
Thus, for sodium cluster anions with up to $57$ atoms the geometrical structures 
have been determined with high accuracy from the comparison of DFT calculations results 
with measured photoelectron spectra \cite{exp,exp1}. 


The measurement of  photoabsorption and photoionization of small mass selected clusters
is not a trivial task. Until recently the measurement was fulfilled only for the 
total photoionization or photoabsorption cross sections of various cluster targets.
In the recent work \cite{Issendorf} the first measurement of angular resolved photoelectron spectra of sodium clusters
has been reported. These experiments have been performed for negatively charged 
sodium ions (anions) in a broad range of cluster sizes  $3\le N\le 147$ \cite{Issendorf}. 
These experiments allowed to probe the angular momenta of single electron orbitals in sodium metal clusters. 
In the work \cite{Issendorf} it was also demonstrated that 
simple models based on single electron treatment of the photoionization process
fail to describe the angular anisotropy of photoelectrons emitted in the process
of photoionization of cluster anions.


In this Letter we present a consistent many-body theory for the description of angular resolved photoelectron 
spectra of metal clusters. As a case of study sodium cluster anions have been chosen.
The Hartree-Fock (HF) approximation has been used as a single particle theoretical framework.
Many-electron correlations have been accounted for within the Random Phase Approximation with Exchange (RPAE).
The results of calculations allow one to conclude that many-electron correlations play the very essential role
in the formation of angular distributions of photoelectrons in the process of photoionization of metal clusters.
This effect is a consequence of the large dynamic polarizability of metal cluster targets, being entirely determined
by many-electron correlations in the vicinity of the  plasmon resonance frequencies.

In this Letter we consider the most characteristic case of a spherically symmetric cluster target and
apply the jellium model for
the description of the photoionization process.
This model proved to be well applicable for the description of electronic structure \cite{Ekardt} and collision processes
involving metal clusters and fullerenes \cite{Solovyov,Solovyov_IJMPB}. Cross sections of the collision processes
are very sensitive to the correct accounting for many-electron correlations, see \cite{Solovyov,Solovyov_IJMPB} and
references therein.  In this Letter we limit our description
by the cluster targets possessing the spherical symmetry, which corresponds
to the case of the so-called magic clusters, i.e. clusters with the entirely closed electronic shells. For the concrete
analysis and illustration we have chosen the $Na_7^-$ and $Na_{19}^-$ magic clusters.

Let us note here also that the angular anisotropy of photoelectrons emitted in the process of
photoionization of atomic negative ions has been investigated in sufficient detail both
theoretically and experimentally (see, for example, review \cite{Ivanov99}). 
It was demonstrated that the parameter of angular anisotropy  $\beta$ \cite{Cooper}
is very sensitive to accounting for the many-electron correlation effects. Thus,
it is not a surprise the failure of single particle approaches reported in \cite{Issendorf}
for the case of the photoionization of cluster targets.



The characteristic features of small metallic clusters, like the shell structure of delocalized  electrons or plasmon 
excitations, can be well understood in terms of quantum motion of the delocalised
valence electrons moving in the field created by themselves and
the positively charged ionic core \cite{Ekardt}. This concept, known as
the jellium model for metallic clusters, can be also utilized for 
the description of a metallic cluster anions.

In this work for the elucidation of the role of many-electron correlations in the formation of 
angular distributions of photoelectrons we have addressed to the most characteristic example. Thus, 
we considered the magic cluster anions possessing the spherical symmetry due to the 
entire closure of  their electronic shells. In this case
the complicated ionic structure can be reduced with the sufficient accuracy to 
the uniformed spherically symmetric positive charge distribution, named below also as ionic core.

Within the Hartree-Fock (HF) approximation the ground state many-electron wave function of the electronic subsystem 
is defined by the Slater 
determinant constructed of single-electron wave functions $\varphi_i(\bf{x})$, being solutions of 
the system of coupled integral-differential equations:
\begin{equation}
(\hat T + \hat U_{core} + \hat V_{el})\varphi_i(\bf{x}) = \epsilon_i\,\varphi_i(\bf{x}) \label{HF},
\end{equation}
where $\hat T$ is the kinetic energy operator, $\hat U_{core}$ is the potential of the ionic core and 
$\hat V_{el}$ is the electron-electron interaction including direct and 
exchange terms. The atomic system of units $|e|=m_e=\hbar=1$ is used throughout the paper.

When calculating the ground state wave functions, the system of equations (\ref{HF}) should be treated 
as coupled and to be solved self-consistently. Each equation in  (\ref{HF}) can be solved by an iterative procedure. 
The eigen-values of these equations are the single-electron energies $\epsilon_i$.
In a spherically symmetric field, the single electron wave functions in (\ref{HF})  can be expressed as 
$\varphi_{n_i,l_i,m_i}(\vec r)$, where $n_i,l_i,m_i$  are the principal, orbital, magnetic quantum 
numbers of the orbital i.
This means that when neglecting the spin-orbital interaction
the cluster electronic shells with quantum numbers ${nl}$ consist of $2(2l+1)$ electrons.
For instance, in the case of  $Na_7^-$ and $Na_{19}^-$ the electronic configurations
of the ground state have the form: $1s^2\, 2p^6$ and $1s^2\,2p^6\,3d^{10}\,2s^2$, respectively.

The wave function of a detached photoelectron can also be obtained
from equation (\ref{HF}) as a solution with the fixed energy $\varepsilon=I_p+\omega$ obeying
certain asymptotic behavior \cite{Amusia1}, 
where $I_p$ is the ionization potential of the initial state and 
$\omega$ is the photon energy. The wave functions of excited states have been calculated
both in the field of a "frozen"\, electronic cluster core with the created vacancy and with
the entirely rearranged  residual electronic structure of the cluster. The latter case corresponds to the situation
of the so-called static rearrangement.




Within the dipole approximation the angular distribution of photoelectrons 
ionized from $nl$-state by linearly polarized light 
can be written in a well-known general form \cite{Cooper}:
\begin{equation}
\frac{d\sigma_{nl}(\omega)}{d\Omega}=\frac{\sigma_{nl}(\omega)}{4\pi}(1+\beta_{nl}(\omega)P_2(\cos\theta)),
\end{equation}
where $\sigma_{nl}(\omega)$ is the total photoionization cross section from $nl$ state, $\omega$ is the photon energy, 
$P_2(\cos\theta)$ is the Legendre polynomial of the second order, $\theta$ is 
the electron emission angle with respect to the polarization of the incident
light. The  angular asymmetry parameter $\beta$ \cite{Cooper} in 
the single-electron approximation reads as \cite{Amusia1}
\begin{eqnarray}
\beta_{nl}(\omega)&=&\frac{1}{(d_{l-1}^2+d_{l+1}^2)}
[(l-1)d_{l-1}^2+(l+2)d_{l+1}^2\nonumber\\
&-& 6\sqrt{l(l+1)}d_{l-1}d_{l+1}\cos(\delta_{l+1}-\delta_{l-1})].
\label{beta}
\end{eqnarray} 

Here $\delta_{l\pm1}$ is the phase of photoelectron, $l$ is the orbital momentum of the ionized shell 
and $d_{l\pm1}$ is the reduced dipole matrix element, which can be obtained from photoionization 
amplitude within the dipole approximation $\langle \epsilon l\pm1 | \hat d | nl \rangle$ 
by the integration over angles and the summation over spin variables. In the length form it
reads as
\begin{equation}
d_{l\pm1}=(-1)^{l_>}\sqrt{l_>}\int\limits_0^\infty P_{nl}(r)rP_{\epsilon l\pm1}(r)dr,
\end{equation}
where $P_{nl}(r)$ and $P_{\epsilon l\pm1}(r)$ are the radial parts of initial and excited states wave functions, 
$l_>=l+1$ for transitions $(l{\to}l+1)$ and $l_>=l$ for the case of $(l{\to}l-1)$.

There is a certain threshold value for the angular anisotropy parameter $\beta$ in the cross section of photodetachment
from negative ions \cite{Ivanov99}. For $l>1$ it is equal to
\begin{equation}
\beta_{nl}\rightarrow \frac{l-1}{2l+1}
\label{lim_beta}
\end{equation}
In the case $l=0$ the $\beta$ parameter is equal to 2.

Expression for the angular distribution of photoelectrons within consistent many-body theory accounting for 
many-electron correlations has been derived for the first time in \cite{Amusia2}. In this formalism
the reduced dipole matrix elements become complex.
The expression for $\beta$ becomes rather cumbersome and reads as \cite{Amusia1}: 
\begin{eqnarray}
&&\beta_{nl}(\omega)=\frac{1}{(|D_{l-1}|^2+|D_{l+1}|^2)}
\{(l-1)|D_{l-1}|^2+\nonumber\\
&+&(l+2)|D_{l+1}|^2 + 6\sqrt{l(l+1)}\times\nonumber\\
&\times&[(ReD_{l-1}ReD_{l+1}+ImD_{l-1}ImD_{l+1})\times \nonumber\\
&\times&\cos(\delta_{l+1}-\delta_{l-1})-\nonumber\\
&-&(ReD_{l-1}ImD_{l+1}-ReD_{l+1}ImD_{l-1}) \times  \nonumber\\
&\times& \sin(\delta_{l+1}-\delta_{l-1})]\}
\label{beta1}
\end{eqnarray} 
where $D_{l\pm1}$ are the complex reduced amplitudes obtained from the dipole matrix elements 
$\langle \epsilon l\pm1 | \hat D(\omega) | nl \rangle$ 
being solutions of the RPAE equation. In 
the matrix form the RPAE equation reads as \cite{Amusia1}:
\begin{eqnarray}
&&\langle \epsilon l\pm1 | \hat D(\omega) | nl \rangle=\langle \epsilon l\pm1 | \hat d | nl \rangle +\nonumber\\
&+&\left(\sum_{i_4>F\atop i_3\le F}-\sum_{i_3>F\atop i_4\le F}\right)
\frac{\langle i_4|\hat D(\omega)|i_3\rangle\langle i_3, \epsilon l\pm1|\hat U|i_4, nl\rangle}{\omega-\epsilon_4
+\epsilon_3+i\delta}
\label{RPAE}
\end{eqnarray}
where indices $i_3$ and $i_4$ describe the virtual electron-hole excitations, 
the combined matrix element $\langle\hat U\rangle$ is a sum of the direct and 
the exchange Coulomb matrix elements $\langle\hat V\rangle$:
\begin{eqnarray}
&&\langle i_3, \epsilon l\pm1|\hat U|i_4, nl\rangle = \nonumber\\
&=&\langle i_3, \epsilon l\pm1|\hat V|i_4, nl\rangle -\langle \epsilon l\pm1, i_3|\hat V|i_4, nl\rangle
\end{eqnarray}

In calculations we take into account correlations between all possible transitions 
$nl\rightarrow \epsilon l\pm1$. 


Using the formalism presented above we have calculated the  angular anisotropy parameter $\beta$
for the photoionization cross section of the magic $Na_7^-$ and $Na_{19}^-$ cluster anions.
Calculations have been performed in the single particle HF approximation and
also with accounting for many-electron correlations.

\begin{figure}[t]
\centering
\unitlength=1cm
\includegraphics[scale=0.85,
clip,angle=0]{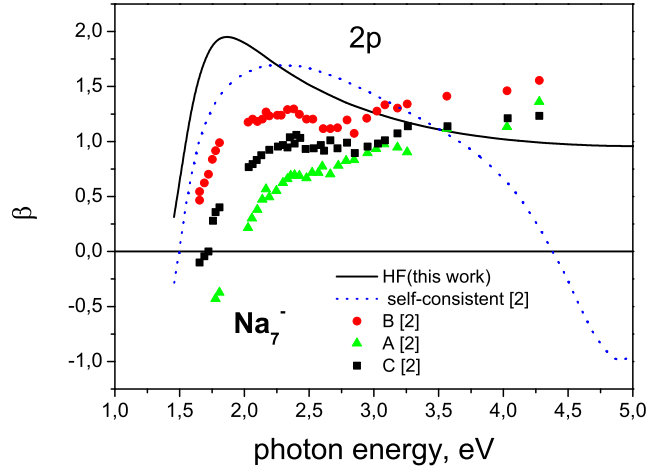}
\vskip -0.2cm
\caption{ Angular anisotropy parameter
$\beta$ for the photoionization  cross section of the $Na_7^-$ cluster 
versus  photon energy; solid line is a result of the HF approach derived in this work;
dotted line is a result of a single-particle approach used in 
\cite{Issendorf}; A, B, C  are the experimentally measured dependencies \cite{Issendorf}.}
\vskip -0.4cm
\label{Fig.1}
\end{figure}

In Fig. \ref{Fig.1} we present comparison of the experimental data from  \cite{Issendorf}
with the results of our calculations of angular anisotropy parameter $\beta$ for
the photoionization of 2p shell of the $Na_7^-$ cluster
performed in the single particle HF approximation.
Different experimental dependencies A, B, C in Fig. \ref{Fig.1} correspond to the photodetachment
of p-electrons from different sublevels of the 2p-orbital arising due to its splitting  by the crystalline field
of the cluster.  In the paper \cite{Issendorf} there was also performed the calculation of
the parameter $\beta$ in a single-particle approximation with the use of the self-consistent Ekardt potential
and additional accounting for
the image-charge and dipole potentials. Our result here is based  on Eq. (\ref{beta}) with
the amplitudes being calculated using the HF wave functions.
As it follows from Eq. (\ref{lim_beta}) the threshold value of $\beta $ for the p-photoelectrons 
should be equal to $0$. From the results obtained it is clear that within a single-particle formalism it is 
neither impossible 
to derive the correct threshold value for $\beta$ nor  the overall correct behavior of $\beta(\omega)$
corresponding to the experimentally measured dependence.

\begin{figure}[t]
\centering
\unitlength=1cm
\includegraphics[scale=0.85,
clip,angle=0]{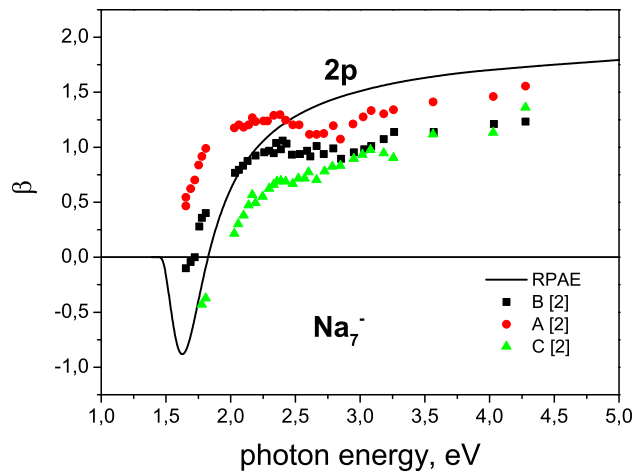}
\vskip -0.2cm
\caption{ Angular anisotropy parameter
 $\beta$ for the photoionization cross section of the $Na_7^-$ cluster versus
photon energy; solid line is a result of the RPAE calculation accounting for
many-electron correlations; A,B,C are the experimentally measured dependencies from  \cite{Issendorf}.}
\vskip -0.1cm
\label{Fig.2}
\end{figure}

In order to elucidate the role of many-electron correlations calculation of the angular anisotropy 
parameter $\beta$  has been performed on the basis of the RPAE equation  Eq. (\ref{beta1}), 
from which the amplitudes $D_{l\pm1}$ have been derived.
In  Fig. \ref{Fig.2} the results of this calculation of $\beta$ for 
the photoionization of $Na_7^-$ cluster are presented and compared with 
the experimental results from \cite{Issendorf}. Comparison of Figs. \ref{Fig.1}  and \ref{Fig.2} 
demonstrates that accounting for many-electron correlations leads to
a qualitative change of behavior of the calculated dependence and thus the angular distribution
of photoelectrons. Fig. \ref{Fig.2} demonstrates also that the consistent many-body theory
leads to the correct threshold value of the parameter $\beta$ and improves substantially
the agreement of theoretical results with experimental data.

\begin{figure}[ht]
\centering
\unitlength=1cm
\includegraphics[scale=0.85,
clip,angle=0]{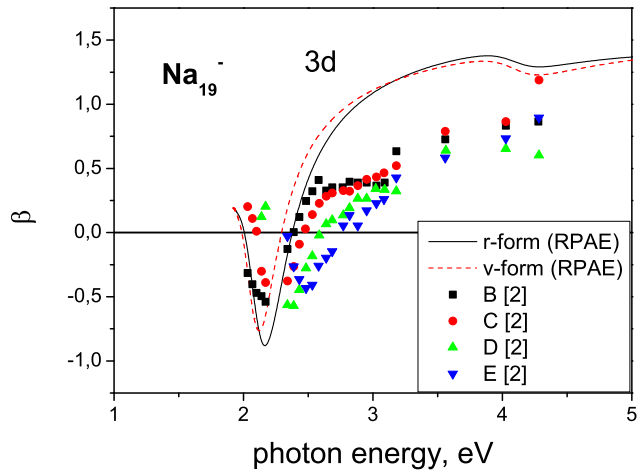}
\vskip -0.2cm
\caption{Angular anisotropy parameter $\beta$ for the photoionization cross section
of the $Na_{19}^-$ cluster versus  energy  of photon;
solid and doted lines are the RPAE results; dependencies  
B,C,D,E are the experimental data  \cite{Issendorf}.}
\vskip -0.5cm
\label{Fig.3} 
\end{figure}

A similar calculation and analysis of $\beta$ have been performed for the photodetachment from the $Na_{19}^-$
cluster anion. In this case  we have calculated
the wave functions for  outgoing photoelectrons in the entirely rearranged electronic cluster
core. In Fig. \ref{Fig.3} we compare the obtained dependence $\beta(\omega)$ for the d-shell
photoionization of the  $Na_{19}^-$ cluster. As in the case of the $Na_7^-$ cluster
different experimental dependencies  B,C,D,E correspond
to the photoionization from the sublevels of the $3d$-orbital splitted by the crystalline field of the cluster. 
Note that the calculated dependence of $\beta(\omega)$ on $\omega$ possesses a characteristic minimum in the
vicinity of the threshold, which is well seen in the all experimental curves.
A small deviation of the parameter $\beta$ calculated using the length and the velocity form 
of the amplitudes $D_{l\pm1}$ arises due to accounting for the electronic cluster core rearrangement
in the static approximation. A small shift in $\omega$ of the calculated curves with respect
to the experimental ones is a result of some deviation of the experimental value of
$I_{p}$ from the one calculated  within the jellium model framework. 

In this paper for the first time the consistent many-body theory has been applied 
to the calculation  of the angular distributions of photoelectrons emitted
the process of photoionization of metal clusters. The concrete analysis
was performed for sodium cluster anions $Na_7^-$ and $Na_{19}^-$ possessing
nearly spherical geometry arising due to the closure of all electronic shells 
of delocalized electrons. For these clusters we have calculated the parameter
of angular anisotropy  $\beta$ of the photoionization cross sections.
These calculations revealed the very important role of many-electron correlations
in the formation of angular distributions of photoelectrons and explained 
the behavior of the angular anisotropy
parameter $\beta(\omega)$  versus  photon energy for sodium clusters recently measured in \cite{Issendorf}.

The next logical step of these investigations could be an extension of the
formalism presented in this Letter for clusters with partially filled shells.
This treatment is to be performed within the framework
of the deformed jellium model \cite{Lyalin}.
The most detail description of the problem, of course, should
take into account the detail geometrical structure of the ionic core.
Also, for the accurate description of the near threshold behavior of the
photoionization cross sections of negative ions one need to account for the effect
of dynamical polarization of the electronic core, i.e. for the polarization of the electronic core by the
outgoing electron, as well as the relaxation processes of the electronic core.
This can be achieved by the method combining the Dyson equation method and RPAE \cite{Ivanov99,Ivanov96}.
All these problems can be addressed for different type of metal clusters (Au, Ag, Mg, Al, K, Sr etc) and
remain open for further investigations in this field.

We are thankful to Dr. Andrei Korol and Dr. Andrei Ipatov for helpful discussions.


\begin{thebibliography}{99}
\bibitem{Issendorf}
C. Bartels, C. Hock, J. Huwer, R. Kuhnen, J. Schwobel, and B. von Issendorff,
Science {\bf 323}, 1323 (2009)

\bibitem{Solovyov}
J.P. Connerade, A.V. Solov'yov (eds), {\it Latest Advances in Atomic Clusters 
Collisions: Structure and Dynamics from the Nuclear to the Biological Scale}, 
Imperial College Press, London, 2008; {\it Latest Advances in Atomic Clusters 
Collisions: fission, fusion, electron, ion and photon impact}, 
Imperial College Press, London, 2004 


\bibitem{Solovyov_IJMPB}
A.V. Solov'yov, 
International Journal of Modern Physics {\bf B19 (28)} 4143 (2005)

\bibitem{exp}
M. Moseler, B. Huber, H. Hakkinen, U. Landman, G. Wrigge, M. Astruc Hoffmann and B. von Issendorff, 
Phys. Rev. B {\bf 68}, 165413 (2003)

\bibitem{exp1}
O. Kostko, B. Huber, M. Moseler and B. von Issendorff, Phys. Rev. Lett. {\bf 98 (4)}, 043401 (2007)


\bibitem{Ivanov99}
V. K. Ivanov, J. Phys. B: At. Mol. Opt. Phys. {\bf 32}, R67 (1999)

\bibitem{Cooper}
J. Cooper and R.N. Zare, J.Chem.Phys. {\bf 48}, 942 (1968); J.Chem.Phys. {\bf 49}, 4252 (1968)

\bibitem{Ekardt}
W. Ekardt (ed.), {\it Metal clusters}, (John Wiley, New York 1999).


\bibitem{Amusia1}
M.Ya. Amisia and L.V. Chernysheva, {\it Automatic system of investigation of atoms}, (Leningrad, Nauka, 1983) 

\bibitem{Amusia2}
M. Ya. Amusia, N.A. Cherepkov and L.V. Chernysheva, Phys. Lett. {\bf 40A}, 15 (1972)

\bibitem{Lyalin}
A.G. Lyalin, S. K. Semenov, A. V. Solov'yov, N. A. Cherepkov and W. Greiner, 
J. Phys. B: At. Mol. Opt. Phys. {\bf 33}, 3653 (2000)

\bibitem{Ivanov96}
V.K. Ivanov, G.Yu. Kashenock, G.F. Gribakin, A.A. Gribakina, 
J.Phys.B: At.Mol.Opt.Phys. {\bf 29}, 2669 (1996) 

\end{thebibliography}
\end{document}